\documentclass[onecolumn]{scrartcl}
\usepackage[utf8]{inputenc}

\usepackage[textwidth=5cm]{todonotes}
\usepackage[left=1cm,right=1cm,top=3cm,bottom=3cm,nohead,nofoot]{geometry}

\title{L'utilisation des données personnelles dans les services publics français : applications, mails, sites internet}
\author{Titouan-Joseph Revol$^a$, Clement Lagneau-Donzelle$^a$, Albert Nicolás López$^a$, Hugo Courté$^a$\thanks{Remerciements à Mathieu Cunche, notre tuteur dans ce projet d'initiation à la recherche (PIR) dans le cadre de nos études à l'INSA Lyon}}
\date{19 Juin 2020}

\usepackage[numbers]{natbib}
\usepackage{microtype}
\usepackage{pgfplots}
\usepackage{graphicx}
\usepackage{float}
\usepackage{hyperref}
\hypersetup{
    colorlinks=true,
    linkcolor=blue,
    filecolor=magenta,      
    urlcolor=cyan,
}

\begin{document}

\maketitle $^a$ INSA Lyon, Département Télécommunications, Projet d'initiation à la recherche

\section{Introduction}

En 2020, les données personnelles sont devenues l’or des multinationales : les GAFAM nourrissent allègrement leur chiffre d’affaire grâce au Big Data, qu’ils revendent ensuite souvent à des entreprises tierces. Toutes les entreprises se battent pour intéresser le plus possible leurs clients en fonction de ce qu’ils aiment manger, voir, faire… pour leur vendre des produits. La collecte des données personnelles est donc clé sur les différents sites web, en témoigne une étude

Cependant face à ces entreprises privées, le service public se doit de faire face en ne cédant pas à ces facilités. Le devoir d'exemplarité de l'État est essentiel pour maintenir la confiance des citoyens. Ils sont pourtant très utilisé d'après une étude de Cookiebot. Cela est d'autant plus dommage que l'Europe a légiféré sur ce sujet avec la RGPD et la loi Informatique et Libertés, ils devraient donner l'exemple sur ce sujet. Une étude \cite{degeling_we_2019} montre d'ailleurs qu'il y a eu des améliorations significatives à l'échelle européenne après la promulgation de cette loi.
En 2019, Cookiebot montrait dans une étude \cite{cookiebot_study_ue} que 89\% des sites officiels de gouvernements de l’UE contenaient des traceurs publicitaires tiers et que 52 entreprises différentes traquaient les citoyens sur le site internet du gouvernement français. Cependant, cette étude n'analyse que les sites web alors que des études \cite{leung_should_2016} ont montré la possibilité notamment sur des applications de pareils problèmes.
Une autre étude \cite{kirch_plusieurs_2015}  menée par la gazette des communes montre que près de la moitié des communes françaises ne sont pas à jour au niveau sécurité, soit 6500 communes, ce qui est très inquiétant concernant la sécurité de nos données. 

Il nous paraissait donc important d'approfondir ces recherches sur la totalité des domaines numériques des services publics français à la fois sur leur site web mais aussi sur les applications ou les mails puisqu'ils sont également un vecteur de récolte de données personnelles, et notre étude montre que beaucoup d'applications du service public font appel à des traceurs douteux, et que certains (rares) mails en contiennent également. En nous focalisant sur l'ensemble des sites webs du service public français, nous avons pu montrer également la présence de grands acteurs tiers via des cookies.

Nous discuterons donc dans la section 2 de nos recherches sur les permissions et traceurs au sein des applications mobiles Android puis dans la section 3, nous verrons si les e-mails font appel à des traceurs en le quantifiant. Enfin, nous observerons ce qu'il en est pour les sites webs de l'ensemble des services publics français avec une couverture plus large que les études citées.

\section{Applications Mobiles}

L'apparition du portable dans notre société a provoqué un nouveau besoin de proximité entre les institutions et les individus grâce aux applications mobiles. Elles sont devenues essentielles dans tous les secteurs, et le service public n'en est pas des moindres. Cependant, bien que ces applications aient une très grande utilité, elles pourraient être aussi un très bon moyen pour tracer ses utilisateurs. Ce traçage semble être comprehensible du point de vue d'une entreprise privée, mais il n'a pas lieu d'apparaître sur les applications concernant les services publics. Pour cette raison, nous avons décidé de faire quelques recherches sur ce domaine.

Pour ce faire, nous nous sommes basés sur le papier \cite{leung_should_2016}, que nous avons essayé de continuer et d'adapter aux services publics. Ceci a été réalisé grâce à certains outils que nous présenterons plus tard. Tout d'abord, nous nous sommes basés sur \href{https://pypi.org/project/play-scraper/}{PlayScraper} pour pouvoir extraire la plupart des applications de PlayStore. Ensuite, nous avons utilisé  \href{https://pypi.org/project/gplaycli/}{GPlayCli} pour procéder au téléchargement des APK correspondant aux applications du service public. Finalement, \href{https://reports.exodus-privacy.eu.org/fr/}{Exodus-Privacy} nous a permis d'avoir les traceurs et permissions associés à chacune de ces applications. Les résultats ont montré une tendance assez claire sur l'utilisation de traceurs dans ce domaine: ils apparaissent dans presque toutes les applications, soit pour faire fonctionner ces dernières, soit pour obtenir des profits grâce à la publicité qui y apparaissent. Par conséquent, sous ces justifications, les services publics ne sont pas l'exception et utilisent aussi des outils pour tracer leurs utilisateurs. 

\subsection{Méthodologie}

La méthodologie à suivre, basée sur les outils que nous avions à notre disposition depuis le début du projet, a été très claire. Nous avons collecté un jeu d'APK des services publics français, basé sur des mots clefs concernant ces derniers, grâce à des critères qui nous ont permis de décider pour chaque application analysée si elle concernait un service public ou pas. Enfin, nous avons analysé ce jeu d'APK grâce à Exodus-Privacy pour avoir les traceurs et permissions associés.

\paragraph{Mots-clefs et critères :}

Tout d'abord, nous avons dû convenir avec des mots qui pouvaient être associés aux services publics pour obtenir les applications associées sur PlayStore. Ces mots, constitués soit par des villes soit par des mots français standards (relatifs à un service public), pouvaient apparaître sur le nom de l'application ou sur le nom du développeur. Ensuite, nous avons dû définir des critères qui nous permettaient de décider pour chacune des applications analysées si elles correspondaient à un service public ou pas. Enfin, nous avons mis ensemble ces deux processus sur un script Python qui les a rendus fonctionnels, de façon à filtrer toutes les applications par développeur, titre, identifiant et site web associé avec l'aide de PlayScraper\footnote{Avec Play Scraper, à cause d’un bug, on ne peut avoir que 50 applications qui sont associées au mot clé considéré, pas plus.}, puis d'agréger ces données dans un fichier. 

\paragraph{Collecte des APKs :}

Dans cette partie, nous avons téléchargé les fichiers APK correspondant aux applications du service public trouvées dans l'étape antérieure grâce à GPlayCli et un script (cf 2.4.2).

\paragraph{Analyse des APKs avec exodus :}

Une fois que nous avions les APKs, nous avons exécuté une analyse locale d'Exodus-Privacy pour connaître les traceurs et les permissions associées à chacune d'elles. Ceci nous a permis d'avoir des résultats que nous avons analysés ultérieurment. 

\subsection{Limitations}
Notre méthodologie a aussi quelques limitations. Tout d'abord, nous n'avons pas considéré les villes avec moins de 5000 habitants, ce qui peut concerner certaines applications de services publics. Ensuite, le script de téléchargement des APK automatisé utilise les API de Google qui semblent avoir un bridage des téléchargements. Certaines applis étaient introuvables ou alors le téléchargement était bloqué et on perdait donc quelques applications. Après, nous nous sommes rendus compte que les analyses Exodus pouvaient varier entre deux itérations. Cela est sûrement dû à des mises à jour desdites applications. Finalement, pour quelques rares applications, l'analyse ne fonctionnait pas en disant que le fichier n'était pas zippé. On a perdu ainsi quelques applications (une dizaine).

\subsection{Critères pour déterminer si une application correspond à un service public}

Nous allons présenter dans cette section les critères que nous avons utilisé pour déterminer si les applications concernaient un service public ou pas : 

\begin{enumerate}

\item Le nom du développeur, concerne-t-il un domaine public?
    
\item Le nom de l’application correspond bien à un possible service public?

\item L’identifiant de l’application contient-elle une extension d’un service public (du genre gouv, nom d'une ville, d'un service public...)?
    
\item L’application nous redirige vers le site web d’un service public? 
    
\end{enumerate}
Si deux de ces critères sont vérifiés, parmi lesquels le 1) et/ou le 4), nous considérons que c’est une application concernant un service public. 

\subsection{Outils utilisés}

\subsubsection{Script d'analyse des applications concernant ou pas les services publics}

Une fois nous avons déterminé les mots-clés qui peuvent être associés aux services publics et nous avons défini les critères qui permettent de décider si une application appartient à un service public ou pas, un script pour analyser puis stocker les applications concernées était de notre besoin. Pour ce faire, nous avons choisi Python comme langage de programmation. Avec l'aide de PlayScraper, qui fournit les méthodes nécessaires pour obtenir les applications reliées à un mot clé et des informations associées, nous avons créé et rempli deux fichiers texte pour stocker les résultats obtenus. 
\\

Sur le premier, nous avons stocké pour chaque application notre décision concernant l'appartenance à un service public ou pas. Une fois le fichier Python est exécuté, pour chacune de ces dernières, un affichage sur le terminal de son nom, son développeur, son identifiant ainsi comme du site web associé est réalisé.  Ensuite, l'utilisateur est invité à une saisie, avec comme possibilités "y" pour oui et "n" pour non, pour chacun des critères exposés sur la section 3.3. Finalement, l'utilisateur doit décider si l'application concerne ou pas un service public à partir de ses réponses antérieures, par une nouvelle saisie qui dans ce cas peut être "y" pour oui, "n" pour non et "t" pour suivant. Le "suivant" est une option permettant de sauter au mot suivant quand l'utilisateur ne veut pas continuer à analyser les applications concernant un certain mot clé, et qui entraîne aussi un "non" pour l'application qui est en train d'être analysée. La décision finale est stockée, avec les réponses précedentes et les 4 paramètres concernant l'application, sur le premier fichier. Sur le deuxième fichier, on stocke juste les identifiants des applications concernant les services publics, ce qui permet aux suivants scripts de procéder au téléchargement de leur APK et à l'analyse de ces dernières.  

\subsubsection{Script de téléchargement automatique des APK}
Il consiste en un outil très simple tenant en quelques lignes et nécessitant juste l'installation préalable de \href{https://pypi.org/project/gplaycli/}{gplaycli}. Dans le dossier /apks il faut éditer le fichier .conf et ajouter les identifiants Google pour le téléchargement. Il parcourt un fichier texte contenant les ID des applications et pour chaque ligne exécute une ligne de commande gplaycli avec pour paramètre l'app-id de l'application. On met aussi en option une barre de progression du téléchargement. Enfin, on affiche le numéro de l'application téléchargée pour permettre à l'utilisateur de voir la progression sur l'exécution totale du programme. 

\subsubsection{Script d'analyse Exodus-Privacy}
Nous sommes passés par deux possibles utilisation d'Exodus. D'abord le projet était de faire un script Selenium pour récupérer les résultats des analyses déjà faites par la plateforme afin de les exploiter. Seulement, le problème était que bon nombre des applications que nous voulions tester n'avaient pas de test Exodus sur le site. L'autre option découverte dans un second temps était d'exécuter les analyses Exodus en local à partir des fichiers APK préalablement téléchargées. C'est donc ce que nous avons fait. Le script développé est assez simple. Il utilise \href{https://github.com/Exodus-Privacy/exodus-standalone}{Exodus Standalone} et pour l'exécuter il faut donc activer l'environnement virtuel (commande : source venv/bin/activate) puis exécuter le script treatment.py. Après exécution, on obtient ainsi un fichier results\_data.csv dans le dossier. Celui-ci contient la synthèse de toutes les analyses ordonnées de manière à obtenir à chaque fois : le nom de l'application, le nombre de permissions demandées, le nom des permissions demandées, le nombre de trackers et le nom de ceux-ci. Le but est ensuite d'exploiter ces résultats de manière à rendre compte du respect ou non de la privacité des utilisateurs dans ces applications.

\subsection{Résultats}
Dans cette partie, nous allons d'abord présenter les résultats sous forme de graphique, qui contiendra les traceurs et les permissions les plus communs concernant l'ensemble des applications. Étant donné que nous avons analysé presque 600 applications concernant les services publics, une analyse individuelle aurait été convenable mais trop large pour notre étude. Pour cette raison, nous allons prendre aussi deux applications réprésentatives de l'ensemble qui contiendront la plupart des traceurs et permissions trouvés pour les analyser plus soigneusement. 

\begin{figure}[H]
\centering
\includegraphics[width=\linewidth]{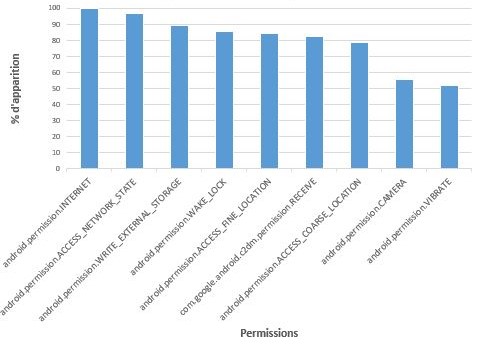}
\caption{Pourcentage d'apparition des permissions demandées dans au moins 300 des 591 applications analysées.}
\label{fig:permissions1}
\end{figure}

\begin{figure}[H]
\centering
\includegraphics[width=\linewidth]{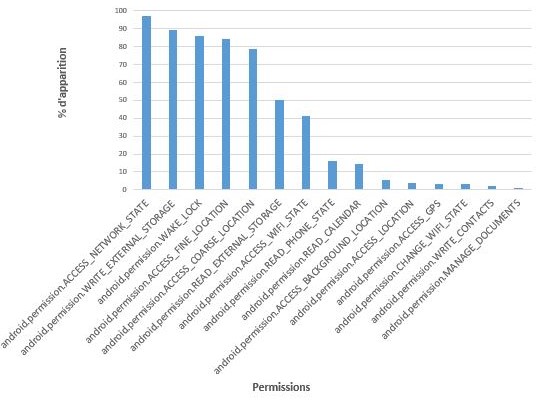}
\caption{Pourcentage d'apparition des 15 permissions les plus intrusives avec la protection des données des utilisateurs.}
\label{fig:permissions2}
\end{figure}

\begin{figure}[H]
\centering
\includegraphics[width=\linewidth]{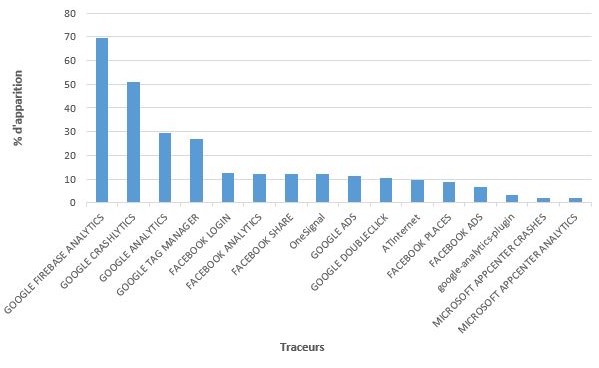}
\caption{Pourcentage d'apparition des traceurs les plus communs sur les 591 applications analysées.}
\label{fig:traceurs}
\end{figure}

\begin{table}[H]
\centering
    \begin{tabular}{|c|c|}
        \hline
            \textbf{Identité} & 
            \textbf{Pourcentage d'apparition} 
            \\
        \hline 
            Google &
            54
            \\
        \hline
            Autres &
            31
            \\
        \hline
            Facebook &
            14
            \\
        \hline
            Microsoft &
            1
            \\
        \hline
                    
    \end{tabular}
    \caption{Pourcentage d'apparition des traceurs selon leur identité}
    \label{table:Traceurs2}
\end{table}

Une autre donnée importante est la moyenne des permissions demandées et des traceurs utilisés dans l'ensemble des applications analysées. Nous avons obtenu une moyenne de 19 permissions et de 4 traceurs par application.  Cependant, pour les traceurs, pour les 45 trackers les plus utilisés, Exodus-Privacy nous donne une moyenne de 3,98 pisteurs par application, ce qui est relativement similaire à celle des services publics. Ils ne font donc ni mieux, ni pire que les autres applications, cependant il paraît primordial que ces applications soient exemplaires.

\subsection{Analyse des résultats}

\subsubsection{Permissions et traceurs}

Tout d'abord, nous allons nous pencher sur les permissions. Sur le graphique présenté en haut (\ref{fig:permissions1}), nous voyons que toutes les permissions qui y apparaissent correspondent à Android. Ceci dénote une claire tendance de la gestion presque totale des permissions par le système d'exploitation concernant le portable en question. Par conséquent, nous pourrions penser que les permissions ne servent qu'à faire fonctionner correctement l'application. Cependant, nous trouvons aussi quelques données significatives. D'abord, nous voyons que presque toutes les applications demandent un accès sur l'état du réseau, ce qui permet de surveiller les connexions sur le réseau. Ensuite, nous trouvons une autre permission qui est très fortement demandée, l'accès à la localisation exacte de l'utilisateur (ACCESS FINE LOCATION). Celle-ci est très significative, étant donné qu'elle permet de contrôler la position exacte de l'utilisateur pendant qu'il utilise l'application concernée, ce qui peut être un utile de traçage très dangereux. De la même manière, nous voyons aussi la permission d'accès à la localisation approximative de l'utilisateur (ACCESS COARSE LOCATION). Finalement, com.google.android.c2dm.permission.RECEIVE donne le permis pour pouvoir enregistrer et recevoir des messages de Google Cloud Messaging, qui est un service qui permet aux développeurs d’envoyer des données depuis des serveurs vers des applications Android.
\\

Maintenant, nous allons procéder à l'analyse du deuxième graphique concernant les permissions (\ref{fig:permissions2}). D'abord, dans ce graphique, nous trouvons quelques permissions déjà analysées avec la Figure 1, comme la localisation ou l'état des réseaux. Cependant, nous voyons aussi l'apparition des permissions WAKE LOCK et WRITE/READ EXTERNAL STORAGE. Ces permissions, utilisées de façon généralisée par les développeurs, sont assez intrusives pour l'utilisateur. La première permet d'éviter que les processeurs du portable s'arrêtent et que l'écran s'éteigne après un certain temps d'inactivité, ce qui peut générer des soucis de batterie. La deuxième et troisième permission sont aussi curieuses, étant donné que sous l'idée de pouvoir écrire ou lire sur la mémoire externe (ce qui peut être nécessaire quand on installe l'application), une possible modification non désirée de cette mémoire ou une réalisation d'une copie de cette dernière peut y être cachée. Ensuite, nous voyons aussi que 40\% des applications demandent l'accès à l'état des réseaux Wi-Fi. Ceci a pour objectif de fournir des données sur ces réseaux Wi-Fi aux développeurs. Après, nous trouvons deux permissions peu communes : READ PHONE STATE et READ CALENDAR. La première permet de connaître l'information contenue dans le dispositif, l'état d'un appel courant et une liste des comptes enregistrés dans l'appareil. La deuxième permet de lire le calendrier de l'utilisateur, pouvant donc permettre de connaître beaucoup d'informations privées et personnelles. Finalement, dans les permissions les moins demandées mais encore qui permettent un traçage de la localisation nous trouvons ACCESS BACKGROUND LOCATION, ACCESS LOCATION et ACCESS GPS. Les trois, très parlantes, additionnées aux deux autres concernant la localisation peuvent être un risque pour le respect de la vie privée des utilisateurs. 
\\

Ensuite, les traceurs vont devenir le centre de notre analyse. Sur la figure \ref{fig:traceurs}, nous avons déjà quelques résultats intéressants à analyser. D'une part, nous voyons que Google est l'acteur préféré par les développeurs dans le choix de leur pisteurs. Les 4 traceurs les plus communs appartiennent à Google, avec Google Firebase Analytics étant le plus utilisé dans la plupart des applications. Ce pisteur permet d'obtenir plusieurs données par rapport à l'usage de l'application par l'utilisateur, ce qui représente une intrusion pas très dangereuse mais considérable de la privacité des utilisateurs. De la même manière, nous trouvons Google Analytics en 3ème position, qui remplit à peu près la même fonction que Google Firebase Analytics. Le 4ème traceur de chez Google, Google Tag Manager, utilisé dans plus ou moins le 30 pourcent des applications, permet de gérer la publicité qui apparaît sur l'application, ce qui constitue encore un outil d'influence des utilisateurs pas négligeable. D'autre part, sur le graphique, nous trouvons des traceurs de Facebook. Ils ne sont pas très utilisés, ce qui est logique étant donné la condition de service public des applications. Nous trouvons aussi deux pisteurs qui n'appartiennent pas à une grande corporation qui sont ATInternet et OneSignal. Le premier a la même fonction que Google Analytics ou Facebook Analytics, et le deuxième permet d'envoyer des notifications aux utilisateurs. Finalement, les pisteurs de Microsoft apparaissent. Dans ce cas, nous constatons qu'ils ne sont presque jamais utilisés. Ceci a une explication très simple : les deux pisteurs que nous trouvons ici, AppCenter Crashes et AppCenter Analytics, remplissent les mêmes fonctions que Google CrashLytics et Google Analytics, donc le choix de ces derniers est préféré pour la grande majorité d'applications concernant les services publics. 

\subsubsection{Identité des traceurs}

Sur la table \ref{table:Traceurs2} nous avons un sommaire de l'identité des traceurs trouvés sur les 591 applications concernant les services publics qui ont été analysées. La première donnée à remarquer, comme a été vu précédemment, est l'appartenance de plus du 50\% du total des pisteurs chez Google. Nous pourrions penser que c'est beaucoup, mais le monopole de l'internet par cette entreprise est largement connu, ce qui provoque que cette entité ait la plus grande dominance sur le secteur. De plus, son pouvoir ne cesse pas d'augmenter grâce à son habilité de stockage des données de tous les utilisateurs du net. Ensuite, nous trouvons une catégorie qui concerne à peu près 1/3 du total des traceurs trouvés. Nous l'avons nommée comme "Autres", étant donné qu'ils appartiennent à des domaines très diverses. Dans cette catégorie nous y trouvons des pisteurs comme ATInternet ou OneSignal, déjà présentés, mais aussi Matomo, Inmobi, MAdvertise, Smart ou Tapjoy, qui sont utilisés soit pour faire fonctionner l'application, soit pour gérer la publicité qui y apparait. Après, nous avons Facebook comme troisième société avec le plus de traceurs. Elle est derrière le 14 pourcent du total des pisteurs, ce qui n'est pas négligeable. Facebook, réseau social le plus influent du monde, utilise la plupart de leurs traceurs pour savoir quel utilisateur est connecté sur son compte dans telle application. Cependant, ces pisteurs sont aussi utilisés pour créer de la publicité ciblée pour les utilisateurs, comme par exemple avec Facebook Ads. Finalement, nous avons Microsoft, qui est présente sur le 1 pourcent des traceurs. C'est une chiffre bas, qui par conséquent confirme la dominance de Google dans le domaine du traçage. 

\subsubsection{Deux applications représentatives : France Bleu et StopCovid}
Pour représenter un peu la présence de trackers et de permissions douteuses, il est intéressant de regarder précisement quelques applications. Ici, nous avons décidé de prendre deux applications représentatives : une application du service public de radio France Bleu de "divertissement" et une application plus confidentielle, et dont les données transitées sont sensibles : StopCovid.

\paragraph{France Bleu}
L'application de la très écoutée radio recèle de surprises.
Du côté des permissions, rien de très surprenant. On retrouve des permissions comme l'accès à Internet, l'affichage au premier plan etc... avec au total 11 permissions (ce qui la classe déjà comme rouge pour Exodus). On retrouve comme permission sensible la géolocalisation précise de l'utilisateur qui sert cependant à la sélection de sa radio (la plus proche) et à l'affichage de la météo.
Cependant du côté des pisteurs, le constat est bien plus délicat. On retrouve en tout 25 "pisteurs" dans l'application dont 5 de Facebook (Ads, Analytics, Login, Places, Share) et 4 de Google (Ads, Crashlytics, DoubleClick, Firebase Analytics). Ici, certains sont relativement peu récolteurs de données comme Login ou Share qui servent à se connecter ou partager des statuts... ou CrashLytics de Google qui sert à signaler les crashs d'applications. En continuant dans les pisteurs plus inconnus, on trouve par exemple AdColony qui sert à l'affichage et la publicité sur des contenus vidéos. Nous avons aussi Adincube, très rare (198 applications référencées sur Exodus soit moins d'1\%) qui utilise l'intelligence artificielle pour augmenter les revenus publicitaires des applications. Nous avons aussi le pisteur Adjust qui s'occupe de cibler les utilisateurs en vue de leur proposer de la publicité ciblée, Millenial Media qui analyse les comportements de l'utilisateurs, récolte des données et affiche de la publicité ciblée. De plus, nous avons Ogury Presage qualifiée sur Exodus comme une plateforme "omnisciente" et qui combine un peu toutes les actions des autres traceurs. Enfin, on trouve des pisteurs plus classiques comme Inmobi, MAdvertise, Smart, Tapjoy, Twitter MoPub et Unity3D qui s'occupent également de la publicité.

\paragraph{StopCovid}
Cette application a fait beaucoup débat à sa sortie : tant au niveau de son principe que de la peur qu'elle piste les utilisateurs. Il est donc très intéressant de regarder son analyse Exodus, d'autant qu'elle est plutôt représentative des applications du gouvernement comme Amendes.gouv et Impots.gouv.
Celle-ci demande 11 permissions ce qui est correct mais qui la classe quand même en rouge sur Exodus. On retrouve comme permissions spéciales l'accès à la gélocalisation précise et à la caméra. Cela reste logique dans le cadre de l'application. Bien qu'elle n'utilise que le bluetooth pour trouver la proximité entre deux utilisateurs est nécessaire pour effectuer un scan bluetooth. La caméra quant à elle est utilisée pour les malades lorsqu'ils doivent scanner un QR code dans l'application pour prouver leur positivité au test. Il faut cependant garder à l'esprit qu'une fois ces permissions accordées, peu de choses empêchent les développeurs de l'utiliser à d'autres fins. Ici, le code source est disponible sur Internet donc vérifiable.
Pour ce qui est de la présence de pisteurs, il n'y en a aucun. En tout cas, aucun pisteur extérieur dans la base de données d'Exodus. Cela demanderait donc plus d'investigations pour en être totalement sûr.

\subsubsection{Types de traceurs : pourquoi ils existent?}
On peut distinguer deux types de pisteurs :
\begin{itemize}
\item Les pisteurs fonctionnels : ils servent à faire fonctionner l'application. On pense notamment à quelques pisteurs comme Crashlytics de chez Google qui remonte les crashs d'applications aux développeurs. On a aussi les pisteurs \href{https://developers.facebook.com/products/facebook-login/}{Login}, \href{https://developers.facebook.com/products/sharing-facebook/}{Share}, ou \href{https://developers.facebook.com/products/places/?locale=fr_FR}{Places} de chez Facebook qui servent respectivement à la connexion grâce à Facebook, au partage sur le réseau social et à l'affichage de lieux sur l'application. Certains "pisteurs" servent à envoyer des notifications à l'utilisateur comme OneSignal, utilisé dans à peu près une application concernant un service public sur 7. Les analytics (Facebook Analytics, Google Analytics, ATInternet ou Matomo qui lui est open-source) servent aussi à analyser comment l'application est utilisée : mesurer l'audience, voir quelles pages sont les plus utilisées... Ils recueillent quand même des données et servent à l'influence de ces acteurs sur l'espace numérique.
\item Les pisteurs publicitaires : ceux-ci peuvent être plus ou moins évolués. Le principe de base est qu'ils s'occupent de proposer les publicités sur l'application dans le but de la rentabiliser voire d'obtenir des profits. Certains se contentent de l'affichage non ciblé de publicités comme Inmobi, MAdvertise, Smart, Tapjoy... d'autres veulent proposer de la publicité ciblée en analysant l'expérience de l'utilisateur. Ils rémunèrent alors les développeurs de l'application en contrepartie. En étant présent sur de multiples applications, ils accroissent leur ciblage des utilisateurs et donc la valeur des publicités qu'ils peuvent afficher. On pense notamment à Millenial Media, Adjust, Adincube...
\end{itemize}

\subsubsection{Pourrait-on s'en passer?}
Il apparaît clair que la présence de ces "pisteurs" est devenue normale. Cependant, le service public se doit de donner l'exemple en protégeant les données de ses citoyens. Pourrait-on simplement les enlever? Dans les applications, certains "pisteurs" sont essentiels : les ressources de réseaux sociaux, la publicité... Cependant ils peuvent être plus ou moins intrusifs et plus ou moins libre et consultables. Les supprimer purement et simplement imposeraient un changement complet de modèle économique pour ceux-ci, vu que le financement par la publicité est classique pour le service public en dehors des applications : les chaînes et les radios y recourent pour assurer le plus possible un équilibre financier. Cependant il paraît primordial que ces traceurs ne soient pas du tout ciblés et c'est bien ici la différence. Le ciblage des utilisateurs est dangereux, surtout pour une institution publique. Il faudrait donc que ces applications contiennent des publicités non ciblées (que l'État mette en place?), ou même qu'il n'y ait plus de publicité du tout et que le modèle économique change.
\\

Les ressources des réseaux sociaux ne sont pas totalement nécessaires. La connexion Facebook Login pourrait être remplacée par France Connect par exemple ou juste une connexion par mail. Le partage sur les réseaux sociaux semble également futile. Les ressources Google Crashlytics sont cependant utiles pour le débogage de l'application.
\\

Pour ce qui est des Analytics, ceux-ci ne sont pas nécessaires mais très utiles aux développeurs pour savoir comment les utilisateurs parcourent l'application. Il paraît possible de s'en passer.

\subsection{Travaux futurs}

Pour les applications mobiles, il serait convenable de faire une analyse encore plus soigneuse qui permettrait de vérifier toutes et chacune des applications du PlayStore/AppStore pour trouver toutes les applications concernant les services publics. Ensuite, on pourrait envisager aussi une analyse individuelle des traceurs et des permissions application par application, qui pourrait détérminer si les services publics derrière chacune de ces dernières respectent la privacité des données des utilisateurs.  

\section{Emails}

\subsection{Méthodologie}
Mon travail s'est inscrit dans la continuité du papier \cite{email_tracking} qui s'est intéressé à la place des traceurs dans les emails en général. Plus particulièrement, j'ai voulu savoir si les emails envoyés par les services publics français étaient exempts de tout traceurs.
Tout d’abord il a fallu définir une méthodologie pour se questionner sur l’approche à adopter, puis réaliser un script qui récupère les données qui nous intéressent. Ensuite il a fallu récupérer des emails, pour en récupérer les URL associées. Ceci m'a permis d’obtenir et d’analyser des acteurs présent dans ces emails.

\subsubsection{Approche}

Premièrement, il a fallu définir quelles étaient les informations à récupérer afin de caractériser si un email contient des traceurs ou non; et comprendre techniquement comment est-ce possible que des trackers soient présent dans les emails. En fait, les emails sont composés de métadonnées et d’un contenu. Ici, les métadonnées nous intéressent très peu, nous récupérons uniquement le domaine de l’expéditeur afin de comparer avec le domaine des URL contenu dans le corps de l’email. Le contenu d’un mail est très souvent en HTML même s’il peut être du text brut. Au début, je récupérais tous les URL dans un mail sauf que le fait est que certains URL sont des liens externes, qui ne sont pas chargés, et qui sont nécessaires dans le corps d'un email. Plus particulièrement, dans ce contenu, les images ainsi que les styles sont du contenu qui est chargé et qui représente une possibilité de récolte de données personnelles. J’ai donc créé un script qui récupère les URL non chargés et les URL chargés dans un emails.
Les emails contiennent souvent deux contenus distincts : un contenu en text brut, qui ne contient par essence aucun traceur, souvent utilisé pour rediriger vers une page web lorsque le contenu en HTML ne peut pas se charger pour un problème technique, et un contenu en HTML. Le traçage est fait lorsque votre utilitaire qui charge votre email HTML charge des images et des styles. Une manière simple de s’en préserver est donc de forcer son utilitaire de mail à ne lire que les text bruts; même s’il est possible de charger du contenu au format HTML, plus agréable pour l’utilisateur, sans qu’il n’y ait de traceur.
Dans la méthodologie, il a également été important de se questionner sur la question des données personnelles, lorsque le script parcours un email, il a accès à tout le contenu de l’email, sachant que certains emails sont assez privés et contiennent des informations sensibles, il a fallu réfléchir à cette problématique.
Le choix qui a été fait est de ne garder uniquement les domaines des sources externes chargés et de ne pas se préoccuper du chemin exact de l’URL d’appel, qui peut, lui, contenir des données personnelles.

\subsection{Présentation}
\subsubsection{Choix des emails}
Deuxièmement, il a fallu caractériser les emails à récupérer. Nous avons choisi de récupérer des emails correspondant aux services publics français tels que l’Assurance Maladie, la CAF, les impôts, le CROUS, EDF, l'ANTS, FranceConnect, des mairies, des factures de cantine ainsi que Parcoursup. Tout cela dans le but de selectionner des mails pertinents. Uniquement des emails provenant de services publics ont donc été traités. Les emails ont été récupérés de manière brut, puis passés au script de récupération des URL chargés.
Pour cela, j’ai pu compter sur l’aide de mes camarades et de Mathieu Cunche, qui m’ont fourni des emails, je les en remercie.
Pour chaque email, je récupère l’ensemble des URL qui apparaissent dans l’email ainsi que l’ensemble des URLS qui sont chargés, à l’aide un parseur HTML et d’expressions régulières.

\subsubsection{Approche par mail}
Pour chaque email ainsi choisi, nous avons décidé de faire un traitement individuel.
Ainsi, dans chaque mail nous récupérons le domaine de l'expéditeur afin de pouvoir selectionner les URL externes.
Pour chaque URL dans le mail, s'il n'est ni du domaine ni w3.org - pour des definitions techniques - on le considère comme externe.

\subsubsection{Problèmes rencontrés}
Lors de la première version du script, les URL chargés ou non étaient traités de la même manière. Ceci était un problème puisque certains URL extérieurs ne sont pas chargés et sont nécessaires dans le corps du mail. Il a donc fallu modifier le script afin de pouvoir séparer les URLS chargés, qui représentent une potentielle fuite de donnée, et les liens externes, qui sont nécessaire dans le corps d'un email.

\subsubsection{Résultats}
J’ai donc analysé un peu plus de 100 emails. Je n’ai trouvé aucun traceur dans la plupart des emails. Ce non-résultat est plutôt intéressant et montre que nos données personnelles sont plutôt bien protégées. Les emails des impôts, de la CAF, d’EDF, de l’ANTS, de FranceConnect sont exempts de traceurs. Je vais vous présenter ici ceux où j’en ai trouvé.
Tout d’abord dans les mails du CROUS, nous retrouvons deux tierces parties, Google et Iroquois.
Dans les mails d’AMELI, de la Sécurité sociale, on retrouve xiti.com.
On retrouve également des traceurs dans les mails de la SMERA, de laposte.net, laposte.fr et dans les emails de la DIRCOM de l’INSA de Lyon.
Plus particulièrement, dans les emails de la DIRCOM de l’INSA, les traceurs de google sont présents.
Les parties prenantes principales que j’ai réussi à identifier sont Google, Iroquois, xiti.com.
On les retrouve également dans les cookies web.

\begin{table}[H]
\centering
    \begin{tabular}{|p{2cm}|p{6cm}|}
        \hline
            \textbf{Acteurs} & 
            \textbf{Présence} 
            \\
        \hline 
            Google &
            CROUS, Smerra, laposte.net, laposte.fr, DIRCOM INSA Lyon
            \\
        \hline
            Xiti &
            Ameli
            \\
        \hline
            Iroquois &
            Crous
            \\
        \hline
                    
    \end{tabular}
    \caption{Présence d'acteurs extérieurs dans les emails}
\end{table}

\subsubsection{Limites méthodologiques}
On ne sait pas si cela est dû à un choix technique et si les acteurs sont pleinement informés du fait que les emails qu’ils envoient contiennent des traceurs.
Enfin, quelques limites méthodologiques, certains URLs considérés comme externe font en fait référence à la même entité. De plus, on sait juste que des traceurs sont présents, on ne sait pas si des données personnelles sont envoyées. Pour cela, le script récupère le lien absolu mais il n’est pas utilisé aujourd’hui pour des questions de respect des données personelles que nous avons évoqué auparavant. Cependant, les questions sur les données personnelles sont beaucoup plus présentes et nécessitent un traitement plus spécifique.

\section{Site Web}

    Dans cette partie, nous allons aborder l'approche avec les sites web. L'idée est de savoir combien de traceurs sont présents sur les sites du domaine public, et de quels types ils sont. Cookiebot avec son rapport \cite{cookiebot_study_ue} datant de 2019 avait déjà essayé de répondre à ces questions. Nous allons baser notre étude sur les cookies. Nous pourrons ensuite comparer nos résultats avec ce rapport \cite{cookiebot_study_ue} pour constater les changements éventuels sur une période d'environ un an. 

    \subsection{Méthodologie}
    
        L’idée est de récupérer ou de constituer une liste de sites web du domaine public. Pour la constitution de cette liste, il faut définir ce qu’est une application publique et les critères qui lui sont associés. Cette définition a déjà pu être faite pour les applications mobiles ou les mails. Nous utiliserons une définition similaire. 
        Cependant sur le portail de l'Open Data du gouvernement, nous pouvons trouver une liste de sites liés au domaine public. Cette liste, disponible à \href{https://www.data.gouv.fr/fr/datasets/liste-des-applications-et-des-versions-mobiles-des-sites-internet-publics/#_}{cette adresse}, nous évite d’en constituer une. Malgré le fait qu’elle date de 2013, elle nous permet d’aller plus vite dans nos recherches. 
        
        Une fois que nous avons constitué une liste, le principe est simple. Il faut visiter chaque site de la liste, dans un premier temps sans aucune action particulière sur le site. Laisser charger la page pour que toutes les ressources soient chargées. À la fin de ce chargement il faut récupérer la liste des cookies liée au site pour ensuite analyser les différents domaines liés. Pour que ce processus soit le plus indépendant possible, pour chaque site visité, les préférences de navigateur doivent être supprimées, ainsi que les cookies.
        
        Ensuite, nous retournons sur le site et acceptons les cookies pour voir les cookies les plus présents. Ceci permet de déterminer si la bannière de mise en garde est vraiment utilisée comme elle doit l'être.
        
        Pour finir, l’analyse de ces données consiste au comptage du nombre de cookies présents sur le site, de relever les domaines associés aux cookies et ainsi faire un classement des sites où il y a le plus de cookies, les domaines qui reviennent le plus souvent.
    
    \subsection{Outils utilisés}
    
        \subsubsection{Récupération des cookies par Selenium}
        
            Pour mettre en place notre méthodologie, nous utilisons \href{https://www.selenium.dev/}{Selenium}. Il permet d’automatiser des tâches sur un navigateur, faire du Crawling ou du Scraping.
            Après avoir paramétré le navigateur pour que celui-ci soit comme à son installation, nous allons sur une des pages de notre liste et attendons que celle-ci charge. Nous pouvons ensuite récupérer les cookies avec un get sur le navigateur. Cependant, nous récupérons que les cookies de session liés au domaine que nous visitons. Pour récupérer les autres cookies, nous allons chercher dans la base de données du navigateur les cookies enregistrés. Là aussi, nous récupérons que les cookies que le navigateur a enregistré. Nous nous retrouvons donc avec normalement la plupart des cookies. Cette méthode a certaines limites qui sont expliquées dans la section “limitations”. 
            
            Par la suite, nous fermons le navigateur, supprimons la base de données et appliquons de nouveau les paramètres par défaut. Ceci permet de simuler une nouvelle visite sur le site. Ce qui évite de charger les cookies qui ont été enregistrés par  notre précédente visite.
            Cette fois-ci, nous cherchons à l’aide de Selenium le bouton d'accord de la bannière d'avertissement des cookies pour cliquer dessus. Une fois le bouton cliqué, nous répétons le procédé d’attente et de récupération des cookies.
    
        \subsubsection{Limitations}

            Avec une récolte des données qui date de 2013, nous nous confrontons à certains risques qui devront être pris en compte lors du développement d’outils. Notamment une partie des sites indiqués peuvent avoir changé d’adresse, qui ne sont plus maintenu ou encore n’existe plus. Pour pallier ce problème, nous faisons d’abord un simple get de la page avec le module requests. En fonction du code de retour HTTP, nous parcourons ou non la page. Nous prenons aussi en compte la non résolution de nom de domaine.
            Cependant, nous ne prenons pas en compte le changement de propriétaire du site en question. C’est à dire qu’avec cette liste, si des sites ont changé de propriétaire et n'appartiennent plus au domaine public, rien n’est fait pour. Il y aura donc dans notre analyse potentiellement des sites qui ne sont pas du domaine public.
            
            Après avoir analysé les résultats obtenus, nous nous sommes rendus compte qu’il manquait des cookies. En effet, nous pensons que les cookies inscrits par des scripts ne sont pas récupérés avec cette méthode. Pour palier à ça nous avons utilisé un autre outil qui peut venir en complément de celui déjà développé. Cet outil est décrit dans la section nouvel outil. 
            De plus, du fait qu’on récupère les cookies à deux endroits différents, nous avons potentiellement des doublons dans nos résultats ce qui veut dire que nous allons compter pour certains sites deux fois les mêmes cookies.
            Nous avons aussi remarqué la présence importante de Google lors de l’analyse de ces résultats surtout dans les résultats de la deuxième visite du site. Cela viendrait-t-il de notre configuration ou du fonctionnement du navigateur ?  Nous avons remarqué la présence récurrente d’un cookie CONSENT de Google. Nous supposons qu’il est lié au fonctionnement du navigateur mais sans réel confirmation car Google Chrome ne parle pas d’une telle fonctionnalité.
            
            L’une des autres limites de méthode est la longueur dûe à la recherche du bouton “accepter” de la bannière. Les méthodes pour afficher cette bannière sont multiples et il n’y a pas de normalisation. Une fois c’est une balise de lien, une autre un bouton. De plus le texte n’est jamais identique (“Accepter”, “Ok, tout accepter”, “Oui, je suis d’accord”, “Ok”, “J’accepte” en sont quelques exemples). 
            La solution que nous avons choisie est un mélange de recherche par valeur du texte, par id et par class. Cela prend donc un certain temps pour trouver le bouton en fonction du site car le script commence par regarder les classes, puis les id puis fait un match avec un xPath.
            
            La dernière limite que nous souhaitons aborder est la parallélisation de notre outil. Il n’est pas possible de le paralléliser sur la même machine car il utilise un navigateur et il y a des conflits d'accès si nous ouvrons une nouvelle instance avec selenium. Et même sans parler de ce conflit, nous aurions un mélange des cookies dans la base de données du navigateur ce qui fausserait nos analyses. La seule solution pour paralléliser notre outil, c’est de découper la liste et de le faire tourner sur plusieurs machines ou machines virtuelles puis de concaténer les résultats pour l’analyse.
    
        \subsubsection{Nouvel outil}
    
            Avec les limites de notre premier outil, nous avons décidé de trouver une autre solution pour récupérer plus de cookies. Nous avons trouvé un outil en ligne nommé \href{https://webprivacycheck.plehn-media.de/en}{Web Privacy Check} qui nous permet de voir les cookies de première party, de troisième party mais aussi les requêtes de troisième party. Cela nous permet d’avoir une idée de tous les cookies et requêtes que le site génère. Nous avons un rapport directement sur l’outil en ligne. Pour récupérer les résultats, nous utilisons selenium pour faire du scraping. Les résultats sont ensuite inscrits dans un premier fichier CSV en tant que donnée brute. C’est à dire que nous récupérons la plupart des informations du rapport, tel que le nom des cookies, le domaine lié, la valeur, le temps d’expiration. Puis dans un second fichier avec un résumé rapide du nombre de cookies de première party et de troisième party et des différents domaines contactés. Ceci permet de simplifier l’analyse plus tard.

        \subsubsection{Limites de cet outil}
    
            Avec cet outil, nous n’avons pas la deuxième partie de notre méthodologie. Le rapport est fait avec une simple visite du site sans action particulière sur celui-ci. Nous ne pourrons donc, avec cette méthode, pas comparer le nombre et le type de cookie avant et après acceptation des cookies.

    \subsection{Résultat}
    
        Dans cette partie, vous retrouverez les résultats basés sur les 6930 sites visités par notre outil.

        \subsubsection{Listes des 10 sites avec le plus de cookies de 3eme partie}
    
            Ici, nous listons les 10 sites qui ont le plus de cookies externes. Nous indiquons le nombre de cookies externes, les domaines associés et les requêtes sur des domaines externes faites.
            
            Avec cette liste nous observons la grande présence de Google mais aussi des GAFAM. Nous pouvons nous rendre compte que les différents sites du domaine public utilisent la solution de facilité offerte par les GAFAM. Nous pouvons remarquer ceci aussi sur le site du gouvernement. La question que nous pouvons nous poser est pourquoi le gouvernement ne met pas à disposition des outils d'analyse et de connexion pour le domaine public. Surtout qu'il existe des outils d'authentification comme FranceConnect qui pourrait remplacer dans de nombreux cas une authentification faite par Google ou par Facebbok.
            
            \onecolumn
            
            \begin{table*}[!hbtp]
                \begin{tabular}{|p{4.2cm}|p{2cm}|p{5cm}|p{5cm}|}
                    \hline
                    \textbf{Site} & 
                    \textbf{Nombre de cookies} &
                    \textbf{Cookies} &
                    \textbf{Requêtes} 
                    \\
                        
                    \hline
                    Cucugnan.fr &
                    33 &
                    Youtube, Tripadvisor, Travel smarter (cookie pour l'optimisation future du site web) &
                    DoubleClick, Authentification Facebook, Google Fonts, Google Analytics  
                    \\
                    
                    \hline
                    Laposte.fr &
                    32 &
                    Doubleclick, Yahoo, 360 yield (une régie pub néerlandaise), Rubicon Project (régie pub américaine), Casalemedia (traceurs d’habitude de navigation), Openx (enregistrement de données géographique) &
                    Google Fonts, Google ads servicies, Amazon ads services 
                    \\
    
                    \hline
                    Gresse-en-vercors.fr &
                    29 &
                    Travel smarter, Tripadvisor, Youtube, Doubleclick, Crwdcntrl (publicité ciblée) &
                    Google Fonts, Google Analytics
                    \\
                    
                    \hline
                    ccimp.com &
                    22 &
                    Doubleclick, Facebook, Twitter, Linkedin, Youtube &
                    Google Fonts, Google Analytics
                    \\
                    
                    \hline
                    fleurat.over-blog.fr &
                    20 &
                    Doubleclick, Facebook, Pinterest, Over-blog, Teads.tv (regie pub), Scorecardresearch (analyse) &
                    Google ads servicies, Authentification Facebook, Google Analytics
                    \\
    
                    \hline
                    lesgauloisdeclemencey.over-blog.com & 
                    18 &
                    Doubleclick, Facebook, Pinterest, Over-blog, Teads.tv, Scorecardresearch &
                    Google ads servicies, Authentification Facebook, Google Analytics, Twitter
                    
                    \\
                    
                    \hline
                    mjc76lillebonne.over-blog.com & 
                    18 &
                    Facebook, Pinterest, Over-blog, Teads.tv, Scorecardresearch &
                    Google ads servicies, Authentification Facebook, Google Analytics, Twitter
                    \\
                    
                    \hline
                    fontenaytorcy.over-blog.com & 
                    17 &
                    Teads.tv, Casalemedia, Rubicon Project &
                    Google ads servicies, Google Fonts, Authentification Facebook, Google Analytics, Twitter, Doubleclick 
                    \\
                    
                    \hline
                    nantesstnazaire.cci.fr &
                    15 & 
                    Linkedin, Dailymotion, Doubleclick, Facebook &
                    Authentification Facebook, Google Fonts
                    \\
                    
                    \hline
                    www.savoie.cci.fr &
                    14 &
                    Linkedin, Doubleclick, Facebook, Youtube &
                    Authentification Facebook, Google Fonts, Google Analytics
                    \\
                    \hline
                    
                \end{tabular}
                \caption{Liste des 10 sites avec le plus de cookies tiers}
            \end{table*}

        \newpage    
        \subsubsection{Les domaines les plus présents}

            Ici, nous avons choisi de montrer les domaines les plus présents dans les sites qui ont été visités par nos outils. Nous retrouvons des grandes compagnies comme le rapport \cite{cookiebot_study_ue} de cookiebot avait pu le montrer.
            
\begin{figure}[H]
                \centering
                \begin{tikzpicture}
                    \begin{axis}[
                        ybar, axis on top,
                        bar width=0.4cm,
                        ymajorgrids, tick align=inside,
                        major grid style={draw=white},
                        enlarge y limits={value=.1,upper},
                        axis x line*=bottom,
                        axis y line*=left,
                        y axis line style={opacity=0},
                        tickwidth=0pt,
                        enlarge x limits=true,
                        legend style={
                            at={(0.5,-0.4)},
                            anchor=north,
                            legend columns=1,
                            /tikz/every even column/.append style={column sep=0.5cm}
                        },
                        symbolic x coords={
                            .google.com,
                            .doubleclick.net,
                            .youtube.com,
                            .xiti.com,
                            .facebook.com
                        },
                       xtick=data,
                       x tick label style={rotate=45,anchor=east},
                       nodes near coords={
                        \pgfmathprintnumber[precision=0]{\pgfplotspointmeta}
                       }
                    ]
                    
                    \addplot [draw=none,fill=blue!30] coordinates {(.google.com, 89)
                                          (.doubleclick.net, 64)
                                          (.youtube.com, 63)
                                          (.xiti.com, 40)
                                          (.facebook.com,4)
                                          };
                                          
                    \addplot [draw=none,fill=red!30] coordinates {(.google.com, 281)
                                          (.doubleclick.net, 468)
                                          (.youtube.com, 351)
                                          (.xiti.com,193)
                                          (.facebook.com,166)
                                          };
                                          
                    \addplot [draw=none,fill=green!30] coordinates {(.google.com, 1435)
                                          (.doubleclick.net, 767)
                                          (.youtube.com, 456)
                                          (.xiti.com,121)
                                          (.facebook.com,704)
                                          };
                    
                    \legend{Number of First party cookies,
                            Number of Third party cookies,
                            Number of Third party requests }
                    
                    \end{axis}
                \end{tikzpicture}
                \caption{Nom de domaine les plus présents}
                \label{Graph_domain_plus_present}
            \end{figure}
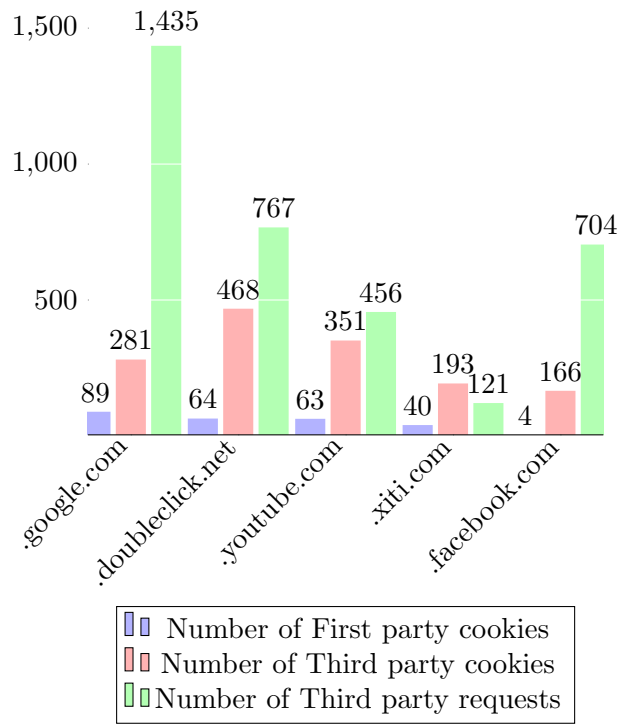

            Le graphique \ref{Graph_domain_plus_present} illustre l'importance des sociétés les plus présentes sur notre échantillon. Nous pouvons voir que cette tête de liste est majoritairement occupée par Google. Facebook est aussi présent. Un autre nom est présent, Xiti, une solution pas très connue du grand public. Elle appartient à AT Internet, une entreprise Française. C'est une solution de mesure d'audience comme Google Analytics mais européenne.
            
    \subsection{Travaux futurs}
    
        Pour le futur, il serait bien de trouver une solution plus efficace pour collecter tous les cookies et requêtes tiers. Cela permettrait aussi de pouvoir faire la comparaison avec l’avant et l'après acceptation des cookies.

\section{Conclusion}

Le traçage des utilisateurs d'internet existe depuis la création de ce dernier. L'information que les entreprises peuvent en tirer est de grande importance pour leurs bénefices, il est donc clair que celles-ci ne vont jamais essayer de se débarrasser de cet outil. Cependant, le service public rentre ici dans un dilemme : doivent-ils utiliser ces traceurs pour obtenir aussi des bénefices ou doivent-ils être exemplaires et faire un traitement respectueux de la vie privée des utilisateurs?

L'étude que nous avons réalisé nous a permis d'extraire quelques conclusions, contrastées avec les résultats obtenus. D'une part, dans le domaine des applications mobiles, les permissions et les traceurs sont presque toujours présents. Android, pour ce qui concerne les permissions de PlayStore, est la principale entité concernant ce domaine. Sous le pretexte de garantir un fonctionnement optimal des applications, ces permissions peuvent quelques fois cacher certains moyens de traçage très dangereux pour les utilisateurs. Google, acteur principal du traçage et puissance majeure en ce qui concerne le stockage d'information d'utilisateurs du net, est derrière la plupart des traceurs. Ceux-ci ont deux missions principales. D'un côté, ils permettent de faire fonctionner l'application, comme les traceurs de Facebook qui servent à se connecter sur son compte dans l'application ou ceux de Google qui permettent soit de remonter des crashs, soit d'analyser comment l'application est utilisée. De l'autre côté, ils permettent de gérer la publicité qui apparaît dans l'application, qui peut être ciblée ou non. Concernant le traçage dans les emails, nous retrouvons des parties prenantes assez présentes : Google, Xiti et Iroquois. Même s'ils sont la plupart du temps utilisés dans le cadre de mesure d'audiance, ils sont présent dans les emails du service publique.
Pour finir avec les sites web, Google est très présent dans les sites du gouvernement. Nous retrouvons des acteurs communs aux applications et aux emails tel que les GAFAM ou encore Xiti. La plupart des cookies présents sont pour des mesures d'audience et de l'affichage publicitaire qui est souvent ciblé.

Maintenant, le défi consiste à réaliser plus de recherches dans ce domaine, analysant plus d'applications, mails ou sites web, puis demandant aux développeurs qui y sont derrière des explications sur le traçage qu'ils réalisent des utilisateurs. Le service public protège-t-il vraiment ses citoyens? Est-il au service de ces derniers ou au service unique et exclusivement de l'État? Ce sont des questions qui n'auraient lieu d'être posées, mais qu'avec les résultats obtenus commencent à prendre du sens...

\bibliographystyle{plain}
\bibliography{references.bib}
\end{document}